\documentclass[]{aa}
\usepackage{psfig}

\begin{document}

\title{X-ray and optical periodicities in X-ray binaries. I.
A0535+26}

\author{V.~Larionov\inst{1,2}
\and
V.M.~Lyuty\inst{3}
\and G.V.~Zaitseva\inst{3}}

\offprints{V.~Larionov}

\institute{Astronomical Institute of St.Petersburg University,
St.Petersburg, Petrodvorets, Bibliotechnaya Pl.~2, 198504, Russia
\and Isaac Newton Institute of Chile, St.Petersburg Branch \and
P.K.Shternberg Astronomical Institute, Moscow, Russia}

\date{Received 16 January 2001 / Accepted 26 July 2001}

\abstract{ A homogeneous set of $UBV$ photometry (354 data points
obtained between 1983 and 1998) for the Be/X-ray binary
\object{A0535+26}~= \object{V725~Tau} is analysed, aiming to look
for possible periodic component(s).  After subtraction of the
long-term variation it was found that only a $\approx 103^d$
periodic component remains in the power spectra in both the $V$
and $B$ colour bands.  The probability of chance occurrence of
such a peak is less than 0.1\%. There are no signs of optical
variability at the X-ray period ($\approx 111^d$). We discuss
possible reasons for a 103-day modulation and suggest that it
corresponds to a beat frequency of the orbital period of the
neutron star and the precession period ($\approx 1400^d$) either
of an accretion disc around the neutron star or a warped decretion
disc around the Be star.  }

\authorrunning{V. Larionov, V.M. Lyuty \and G.V. Zaitseva}

\titlerunning{Periodicities in XRB:  A0535+26}

\maketitle

\section{Introduction}

The X-ray binary \object{A0535+26} has been an object of interest
to both observers and theorists for a quarter of a century, from
the moment of its first documented X-ray outburst in 1975
(Rosenberg et al.~\cite{Ros75}). As soon as an optical counterpart
was identified almost simultaneously by several authors
(Liller~\cite{Lil75}; Murdin~\cite{Mur75}) with the 9th magnitude
Be star \object{HDE~245770}, later named \object{V725~Tau}, there
were several attempts to find the orbital period of the system,
using both X-ray and optical (spectral and photometric) data.

Priedhorsky \& Terrell (\cite{PT83}) have shown that all but the
most powerful X-ray outbursts (1975 April and 1980 October) take
place with a period $\approx$111 days. Now the general opinion is
that the orbital period is 110--111$^d$. The most reliable
determination of the X-ray period and initial epoch is currently
that of Motch et al.  (\cite{Mo91}): $\mbox{P}_\mathrm{XR} =
111\fd 38\pm 0\fd 11$ and $T_0 = \mbox{JD} 2446734.3\pm 2\fd 6$.

This and other periodicities were claimed to have been found in
optical (photometric and spectral) data. For instance, Hutchings
et al. (\cite{Hu78}) found that the radial velocities of
absorption lines of \object{HDE~245770} are variable, and
suggested several probable periods:  28.6, 48 and 94 days. Some
years later and based on a very similar dataset, Hutchings
(\cite{Hu84}) found a 112 day period.  Guarnieri et al.
(\cite{Gu82}) noted at first possible 32, 63 or 77 day periods in
their photometric data, but later (Guarnieri et al. \cite{Gu85})
found modulation in the $V$ band photometry with the proposed
orbital period of 110 days.  Gnedin et al. (\cite{Gn88}) made a
Fourier analysis of $V$ band photometry obtained during 1981--1985
and marked out periods of 1100, 103 and 28 days, but found no
periodicity corresponding to the X-ray period.

Besides the determinations noted above, there are several more
publications on this subject, listed by Giovannelli and Graziati
(\cite{GG92}).

In a recent paper by Hao et al. (\cite{Hao96}) all the published
data sets from 1981 to 1993 have been analysed together. Their
analysis have shown only two significant periods -- 830 and 507
days. However their light curve modeling does not agree with the
observational data and their prognosis of the photometric behavior
of \object{HDE~245570} for 1994--1997 was not confirmed by later
observations (see Fig.~1 in Hao et al.  \cite{Hao96}, Fig.~1 in
Lyuty \& Zaitseva \cite{LZ2000} and Fig.~\ref{fig:2} in this
work). It should be noted also that in Hao et al.~(\cite{Hao96})
no correction has been made for different zero-points and colour
systems of the data sampled by different groups.

As soon as our data set had grown substantially and surpassed that
used in Gnedin et al. (\cite{Gn88}), as well as other published
data sets, we attempted to search for periodic components in the
light curves of \object{A0535+26}/\object{V725~Tau}. In Sect.~2 we
describe our techniques of data analysis and obtain a value of the
optical period, distinctly different from the orbital one, while
in Sect.~3 we discuss possible reasons of the light modulation and
suggest that it is caused by interplay of the orbital and
precession motions. In Sect.~4 we summarize the results.

\section{An analysis of optical data}

All the photometric data discussed herein have been obtained in
the period 1983--1998 with a photon-counting photometer at the
60~cm telescope of the GAISh Crimean laboratory (for details see
Lyuty \& Zaitseva \cite{LZ2000}). More than 350 individual $UBV$
measurements have been made; the photometric accuracy is $ 0\fm
005$ in $B$, $V$ and $0\fm 008$ in $U$. As the long-term behavior
of \object{V725~Tau} has been discussed in separate papers (Clark
et al. \cite{C99}, Lyuty \& Zaitseva \cite{LZ2000}), we restrict
ourselves here only to the analysis of variability on the time
scales close to the X-ray period.

\begin{figure}[tbh]
\psfig{figure=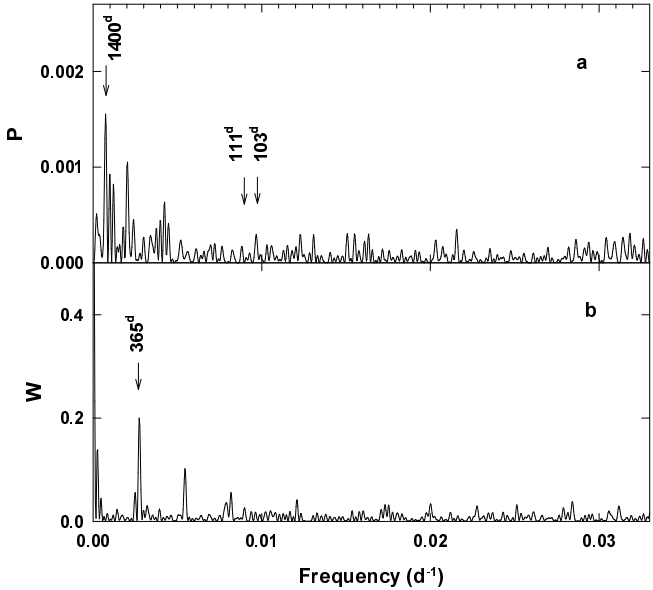,width=88mm,clip=}
\caption{{\bf a} Power spectrum and
{\bf b} spectral window in the $V$ band for our 1983--1998
observations after subtraction of the linear trend. The highest
peak corresponds to a (quasi)period of $\approx 1400^d$.}
\label{fig:1}
\end{figure}

We calculate the power spectrum and spectral window of the data
set as

\begin{equation}
P(f)=\frac{1}{N^2}\left | \sum\limits_{j=1}^N
(m_j-\overline{m})\mathrm e^{-\mathrm{i} 2\pi f t_j}\right |^2
\label{eq:ps} \end{equation}

and

\begin{equation}
W(f)=\frac{1}{N^2}\left | \sum\limits_{j=1}^N
\mathrm e^{-\mathrm{i} 2\pi f t_j}\right |^2, \quad
W(0)=1, \label{eq:w} \end{equation}

\noindent respectively, where $\mathrm{i}=\sqrt{-1}$,  $t_j$
denotes the time of observation, and $m_j$ and $\overline{m}$,
correspondingly, the individual and mean values of the brightness.

\begin{figure}[tbh]
\psfig{figure=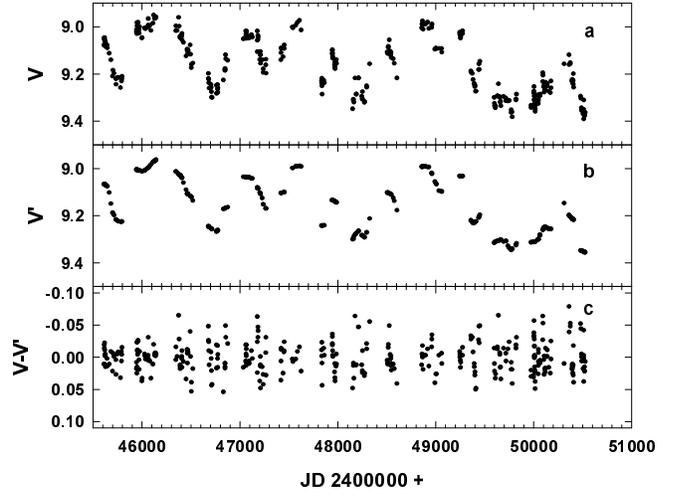,width=88mm,clip=}
\caption{{\bf a}
The light curve of V725 Tau in $V$ band, {\bf b} smoothed light
curve with a 50$^\mathrm{d}$ window, and {\bf c} residuals after
subtraction of the smoothed curve.} \label{fig:2}
\end{figure}

Fig.~\ref{fig:1}a shows the power spectrum of the $V$ band data
(linear trend removed) for the period from 1983 to 1998. The
largest peak in the spectrum corresponds to $\approx 1400$ days
and is readily explainable: a wave with that characteristic time
scale is easily seen in Fig.~\ref{fig:2}, superimposed on a smooth
decay in brightness, in the optical and infrared light curves (see
also Larionov \cite{L93} and Clark et al. \cite{C99}). In the
region of the orbital period ($\sim 0.01~\mathrm{d}^{-1}$), only a
faint peak corresponding to $P=103^\mathrm{d}$ is seen. At the
frequency corresponding to a 111 day period, there is no peak at
all.

\begin{figure}[tbh]
\psfig{figure=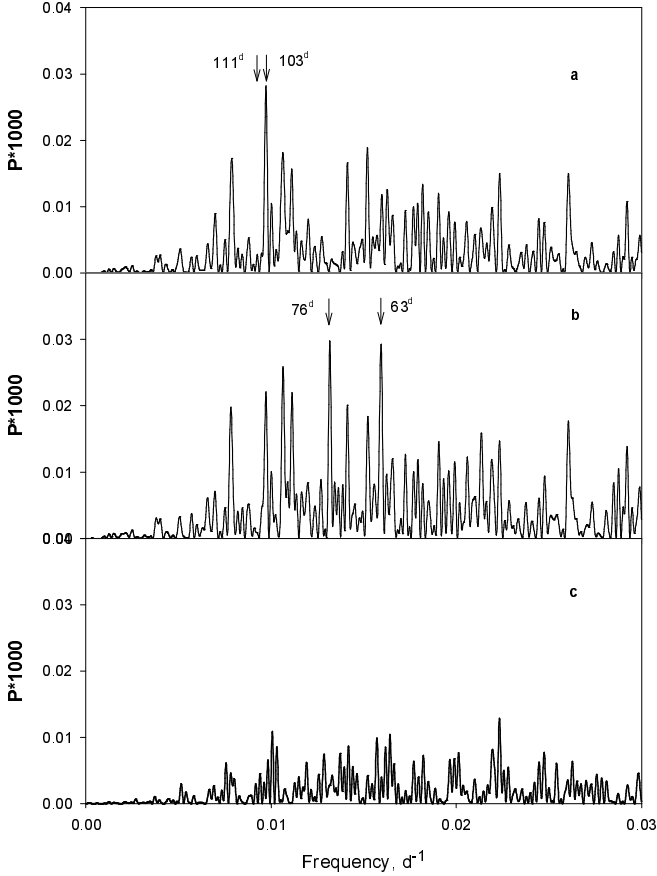,width=88mm,clip=}
\caption{{\bf a}
Power spectrum of the residuals $V-V^\prime$ on our 1983--1998
data. The most prominent peak corresponds to $P=102\fd 92$. {\bf
b} To check the efficiency of smoothing, an artificial sine-wave
with $P=76^\mathrm{d}$ amplitude $0\fm 015$ was added. Additional
peaks in the spectrum, the most prominent of which are at
$P=63^\mathrm{d}$, are caused by aliasing of the $P=76^\mathrm{d}$
period with the peaks in the spectral window. The slight decrease
in the power of $P=102\fd 83$ peak, as compared to the upper
panel, is caused by a redistribution of power as a result of the
cleaning process on our non-uniformly spaced data set. {\bf c}
Power spectrum after subtraction of $102\fd 83$ and $91\fd 07$
period sine-waves from the initial light curve.} \label{fig:3}
\end{figure}

In order to judge with sufficient confidence the reality of the
orbital plus close and/or related periodicities, one should remove
the longer-period component(s) of variability. In our case it is
not sufficient to subtract the linear trend, and we need to take
into account the slow light variations with a (quasi)period of
$\approx 1400$ days. We have constructed an approximating data set
using the method of a sliding mean with the window value $\Delta
$, replacing raw data $m_i$ for each time $t_i$ by the weighted
mean:

\begin{equation}
m_i^{\prime}= -2.5\log\left(\frac 1{\sum p_j}
\sum\limits_{j=1}^k p_j \cdot 10^{-0.4m_j}\right), \label{eq:1}
\end{equation}

\noindent where $k$ is the number of data points within interval
$ [t_i-\Delta, t_i+\Delta]$, and the weight of $j$th point is
determined as

\begin{equation}
p_j=\exp\left[-(\delta t_j/\Delta)^2\right],
\label{eq:2}
\end{equation}

\noindent where $\delta t_j$ is the time span from the $j$th point
to the center of the window. The optimal value of the smoothing
interval was searched by trial and error within a range
$20^\mathrm{d}< \Delta < 100^\mathrm{d}$. The criterion for
$\Delta$ selection was the signal-to-noise ratio of the
power-spectrum peaks in the region of interest, i.e. around the
orbital frequency. The raw and smoothed light curves, and also the
residuals between them, are plotted in Fig.~\ref{fig:2}.  The
power spectrum for the residuals $m_j-m_j^{\prime}$ is shown in
Fig.~\ref{fig:3}.  We have found that, independent of the value of
the smoothing interval adopted, the most prominent peak in the
power spectrum corresponds to a period of $102\fd 83 \pm 0\fd 1$,
while the low-frequency components are most effectively suppressed
and the highest signal-to-noise ratio is obtained when $\Delta =
50^\mathrm{d}$. We have obtained an analogous result for the $B$
band as well.

The amplitude of the sine-wave obtained is $0\fm 015$ in $V$ and
$0\fm 01$ in $B$, while in $U$ the 103-day peak is not seen above
the noise. The significance of the detection of the faint
sine-wave superimposed on the large-amplitude slow variability can
be tested by adding an artificial low-amplitude harmonic to the
raw data. We added a sine-wave with $P=76^\mathrm{d}$ and
amplitude $0\fm 015$ to the raw $V$ data, then repeated the same
cleaning routine as described above (Eqs.~\ref{eq:1}
and~\ref{eq:2}). Fig.~\ref{fig:3} demonstrates that the
low-frequency filtration method described above can be used to
detect, {\it at least for the data set discussed}, periodic
variability on a time scale of $\sim 100^\mathrm{d}$ with
amplitude $\ge 0\fm 01$.

\begin{figure}[tbh]
\psfig{figure=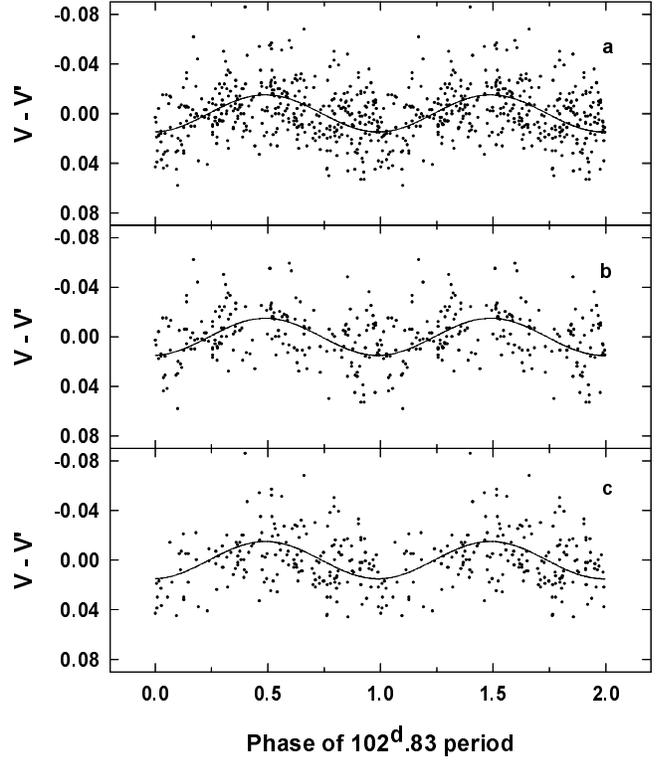,width=88mm,clip=}
\caption{{\bf a} $V-V^\prime$ dependence on the phase of the
103-day period for the whole set of data, {\bf b} for the
``upper'' and {\bf c} for the ``lower''  part of it. A 0\fm 015
sine-wave is superimposed on each panel.} \label{fig:4}
\end{figure}

In order to test whether the 103-day period is real or an artifact
intrinsic to some part of the data set, we have split the initial
data set into two parts corresponding to values above and below
the linear trend and have repeated the same procedure. We have
found that the 103-day harmonic is intrinsic to both parts of the
initial dataset, although for the low-brightness part the
significance of the $103^\mathrm{d}$ peak is slightly less.
Fig.~\ref{fig:4} gives the phase curves of residuals $V-V^\prime$
for the whole data set and for the ``upper'' and ``lower'' parts
of it. It is easily seen that both the shape and initial phases
are the same. We calculate the ephemeris of the small-scale
optical variations as $\mbox{JD}_\mathrm{Min} = 2448002(\pm 3\fd
0) + 102\fd 83(\pm 0\fd 15)\cdot E$, where $E$ is the epoch
number.

The probability of chance occurrence of a peak with amplitude
$P_\mathrm{max}$ in the power spectrum with a mean value
$P_\mathrm{mean}$  can be estimated as $\Phi =100\%\cdot
\{1-[1-\exp
[-(P_\mathrm{max}/P_\mathrm{mean})]]^{N_\mathrm{ind}}\}$; for
non-uniformly spaced data  the number of independent frequencies
$N_\mathrm{ind}=-6.362+1.193 \cdot N+0.00098 \cdot N^{2}$, where
$N$ is the number of observations (Horne and
Baliunas~\cite{HB1986}). In our case $N=354$ and
$N_\mathrm{ind}=539$; after subtraction of the slow component as
described above (Eqs.~\ref{eq:1} and \ref{eq:2}), $\Phi <0.1\%$.

Nevertheless, there are some peaks in the frequency range
$0.007-0.017$ with power $>10^{-5}$. In order to test their
reality, we have followed the recipe given by Horne and
Baliunas~(\cite{HB1986}) for irregularly spaced data and
subtracted a $103^\mathrm d$ period sine-wave from the initial
light curve. After this we find that only a peak at a frequency
corresponding to $91\fd 07$ and its aliases remain in the power
spectrum. This feature replaces the $103^\mathrm d$ periodicity
from $JD~2449400$ (1993 April) onwards. It is remarkable that this
change of periods coincides with the time of the most prominent
X-ray outburst over the past three decades.

Fig.~3c displays power spectrum after subtraction of a sine-wave
of period $103^\mathrm d$ before 1993 April and $91^\mathrm d$
after that time. Comparing Fig.~3a and Fig.~3c, we conclude that
only the $103^\mathrm d$ and $91^\mathrm d$ peaks are real,
although the latter feature is present in only $1/5$ of the total
data set, thus preventing us from making exact estimates of the
ephemeris. A slight enhancement of the noise level around
$0.01$~d$^{-1}$ on Fig.~3c is most probably caused by the
deviation of the real signal from a true sine-wave.

\section{Discussion}

We stress once again that the 111-day orbital periodicity is not
revealed in the optical photometry. It is then natural to suppose
that one of the constituents of the total radiation of the system,
besides that of the optical star, is that of a precessing disc --
either an accretion disc around the neutron star or a
tilted/warped equatorial envelope around the optical star. The
period of precession, whichever precessing body, would then be
$P_\mathrm{prec}=1/(1/P_\mathrm{opt}-1/P_\mathrm{x-ray})=
1360^\mathrm{d}\pm 30\mathrm{d}$, and $\sim900^d$ after 1993
April. In the following we consider these two scenarios
separately.

\subsection{Accretion disc}

In the case of the neutron star (NS) accretion disc, the axis of
the disc is inclined to the orbital plane of the system and,
additionally, is counter-precessing with a $\approx
1360^\mathrm{d}$ period. In this case the disc cross-section, as
seen from the optical companion, would be changing with a
frequency equal to the sum of orbital and precession frequencies,
which would lead to the 103-day modulation observed.

The existence of an accretion disc in this binary system at least
during some X-ray outbursts is firmly established -- it is
confirmed by quasi-periodic oscillations of X-rays during X-ray
outbursts (Finger et al. \cite{F96},  \cite{F96}) and by neutron
star spin-up episodes (Nagase et al. \cite{N84}). However, the
question as to whether the disc exists permanently or is formed
just before the outburst remains unresolved.

A similar model was proposed to describe the optical light curves
of some other X-ray binaries: \object{Cen~X-3}, \object{LMC~X-4}
and others (see, e.g.,  Heemskerk \& van Paradijs \cite{HvP89}) --
however, unlike \object{A0535+26}, these systems show orbital
variability besides precession. This can be explained by the fact
that \object{A0535+26} is a wide binary, and the effects caused by
the ellipsoidal shape of the optical component and/or its X-ray
heating are unobservable. Also, in the case of \object{A0535+26}
the large eccentricity of the orbit -- $e\approx 0.5$ (Finger et
al. \cite{F94}, \cite{F96}) -- plays a major role.  As a result,
both the effective cross-section of the disc to the stellar wind
and its illumination in periastron and apastron would differ by
$\sim 9$ times; when $\approx 1360$ day precession is added, this
enables the 103-days modulation to be observed.

Meanwhile, the disc's input to the total optical radiation of the
system can be as small as a few per cent (Lyuty \& Zaitseva
\cite{LZ2000}), and it would be quite a difficult task to
distinguish it based on the difference between its photometric
and/or spectral parameters and those of a giant star and its
envelope.

The disc is not expected to rotate as a solid body, and its
structure is dependent on X-ray outbursts occurring at the surface
of the neutron star and subsequent re-radiation of stored energy.
Because of this, neither the amplitude of precession modulation
nor the period and phase of precession should be stable.

\begin{figure}[tbh]
\psfig{figure=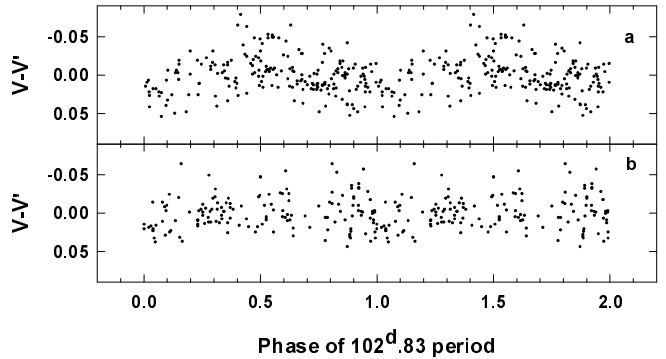,width=88mm,clip=}
\caption{$V-V^\prime$ dependence on the phase of the 103-day
period {\bf a} for descending and {\bf b} ascending  branches of
the slow component of the light curve (see Fig.~\ref{fig:2}).}
\label{fig:5}
\end{figure}

Meanwhile, we see that during the first 10 years of our
observations (up to 1993), the 103-day periodicity of the optical
variations is retained, which means that the precession period is
rather stable, changing from $\sim1360^\mathrm d$ to
$\sim900^\mathrm d$ after the major X-ray outburst. Nevertheless,
as expected, in the total radiation of the system the relative
contribution of the component connected with the disc precession
is not constant. Fig.~\ref{fig:5} illustrates the phase curves in
$V$ band after subtraction of the slow component, separately for
the ascending and descending branches of the initial light curve
shown in Fig.~\ref{fig:2}.  It is evident that the 103-day
modulation is practically absent when the light level is
increasing, and, on the contrary, is quite distinct when the light
fades.

One of the possible mechanisms of stimulation of the accretion
disc precession is free precession of the neutron star. It was
shown by Schwarzenberg-Czerny (\cite{Sch92}) that
$P_\mathrm{pr}/P_\mathrm{spin}\approx 10^6$, where $P_\mathrm{pr}$
is the period of free precession of a neutron star and
$P_\mathrm{spin} $ is the period of its axial rotation. In the
case of \object{A0535+26} $ P_\mathrm{spin}\approx 104^\mathrm s$
and correspondingly $P_\mathrm{pr}\approx 1200^\mathrm{d}$. It is
natural to suppose that the proximity of this period to the
characteristic time of the large-scale photometric variability is
not coincidental. What then can serve as a ``transmission link''
from the freely precessing neutron star and its accretion disc to
the equatorial envelope of the optical component? Let us be
reminded that the neutron star in A0535+26 system has a powerful
magnetic field, $\approx 10^{13}$~G. Periodic changes in
orientation of the magnetic field relative to the equatorial
envelope would lead to periodic changes in the strength of their
interaction, first of all near periastron; the shock wave arising
in the envelope would lead to (quasi)periodic ejections of the
envelope. Clark et al. (\cite{C99}) and Lyuty \& Zaitseva
(\cite{LZ2000}) argue that the long time scale variability of
A0535+26 in the period we are analysing is explained by the
successive expulsions of the Be star envelope that occur on a
characteristic time-scale of $\approx 1400$ days. We have to note,
however, that the detailed description of the interaction of the
Be star envelope and NS magnetic field needs additional modelling
work that is beyond the scope of this paper.

\subsection{Be-star equatorial envelope}

A quite different approach may be considered on the basis of
recent analysis of spectral variability of another Be/X-ray
binary, \object{V635~Cas}=\object{4U0115+63}, made by Negueruela
and Okazaki (\cite{NO2000}) and Negueruela et al.
(\cite{Neg2001}). These authors argue that in \object{V635~Cas},
just as in similar Be/X-ray systems, a major role is played by an
equatorial decretion disc around the optical star. This disc is
truncated as a result of tidal/resonant interaction with the
neutron star companion. At a certain stage the disc becomes
unstable, tilts and warps and starts to precess. Later on, the
disc is disrupted due to interaction with the orbiting neutron
star, and a giant X-ray outburst may occur. The above authors
argue that this model does not imply any substantial change in the
optical star's mass loss rate.

Following that idea, we may suppose that the precessing body in
the \object{A0535+26} system is a tilted/warped disc around Be
star. Then it is natural that the gravitational pull of the
neutron star in the moments of closest approach of the NS to the
disc causes its distortion, which is observed in the light curve
as minor variations superposed on global changes. Due to
precession, the times of closest approach do not coincide with the
periastron passages, but rather precede them by
$\approx8^\mathrm{d}$ after each orbit. Within this approach we
can find a natural explanation for the fact that during the
ascending parts of the global light curve, no 103-day modulation
is seen: it only means that the decretion disc is not large enough
to become tilted and warped and therefore it does not precess.

To further develop this model, one may speculate that the changes
of projection of the warped disc to the plane of sky, as seen from
the optical companion, should cause substantial variations with
the precessional period, superposed on the optical light curve.
This is actually observed: we noted before the coincidence of the
$1400^\mathrm{d}$ time-scale of global optical variations with
almost the same period of precession -- whichever the precessing
body.

One might expect that the X-ray activity within the system may
also in some way be connected with the phases of the
$103^\mathrm{d}$ period. We tried to check whether such a
correlation exists, and found that out of 18 documented outbursts
with amplitude $\geq 0.2$~Crab (Giovannelli \& Graziati,
\cite{GG92} and Finger et al. \cite{F96}), 7 peaked in a narrow
interval $0.6-0.7$ of the optical phase curve (Figs.~\ref{fig:4}
and~\ref{fig:5}), where phase 0.0 refers to optical minimum (see
Table~\ref{tab:1}). This means that X-ray outbursts ``prefer'' to
happen 10 to 20 days after the optical maxima.  If we consider
this in the context of the warped decretion disc model, this delay
can be explained as the time needed for the matter captured from
the Be-star disc -- disturbed during the close passage of the
neutron star -- to travel to the vicinity of the NS companion,
where it loses its angular momentum, falls onto the NS, and causes
an X-ray flare. It can be noted that during 2001 two occasions of
favourable X-ray and optical phases will be in June and October;
unluckily, the former date corresponds to the seasonal gap of
observations.

\begin{table}
  \centering
\begin{tabular}{|c|c|c|}
\hline JD2400000+ & X-ray flux & $\Phi_\mathrm {102\fd8}$\\ &(Crab
units)&\\ \hline

42533 &   2.8 &   0.81 \\ 42614 &   0.2 &   0.60 \\ 42724 &   0.3
&   0.67 \\ 42829 &   0.3 &   0.69 \\ 43288 &   0.5 &   0.15 \\
43508 &   0.5 &   0.29 \\ 43617 &   0.2 &   0.35 \\ 43732 &   0.7
&   0.47 \\ 43951 &   0.2 &   0.60 \\ 44522 &   1.5 &   0.15 \\
44952 &   0.2 &   0.33 \\ 45290 &   0.2 &   0.62 \\ 45515 &   0.5
&   0.81 \\ 45619 &   0.8 &   0.82 \\ 45732 &   0.8 &   0.92 \\
46736 &   0.8 &   0.68 \\ 47625 &   0.6 &   0.33 \\ 49403 &   1.4
&   0.62 \\

 \hline
 \end{tabular}
  \caption{Julian dates, X-ray fluxes and optical phases of
the outbursts with $F(2-10)\mbox{keV}\geq 0.2$~Crab. X-ray data
are from Giovannelli \& Graziati (1992) and Finger et al. (1996).
Giant outbursts of 1975 and 1980 are included.} \label{tab:1}
\end{table}

\section{Conclusion}

An analysis of the uniform photometric data set obtained in the
period 1983--1998 has allowed a confident separation of the
periodic constituent in the light curve of the high-mass X-ray
binary, A0535+26/V725~Tau. The parameters of this periodic
component and its link with the phase of activity of the optical
component allow us to suggest precession of an accretion disc
around the neutron star or a warped equatorial disc around a Be
star as the most likely mechanisms. At this point, both models
seem to be viable; meanwhile, the analysis of already existing
spectral data could be helpful, in the sense that if the warped
disc model does reflect reality, one might expect to see $V/R$ and
$EW$ variations corresponding to the precessional motion.

Within both models, we do not expect that substantial X-ray
outbursts to occur during ascending parts of the large-scale
optical light curve. Moreover, X-ray outbursts tend to occur at
specific phases of the $102\fd 8$ optical light curve. Taken
together, both effects can explain the ``missing outburst''
phenomenon.

Our results lead us to suppose that in other similar systems we
can hope to distinguish the minor photometric variations close to,
but not necessarily coincident with, the binary rotation period.
In future papers of this series, we plan to apply the techniques
described here to the analysis of the light curves of similar
systems, such as \object{X~Per} and \object{V635~Cas}.

\begin{acknowledgements}
V.Larionov acknowledges support from the Russian Federal Program
``Integration'' (project K0232). We thank Catherine Brocksopp and
Alan Marscher for critically reading the manuscript and for
helpful discussion.
\end{acknowledgements}

\bibliographystyle{apj}

\end{document}